\documentclass[twocolumn,floatfix,superscriptaddress,a4paper,showkeys,nofootinbib,reprint]{revtex4-1}

\usepackage[colorlinks=true,linktocpage=true,linkcolor=blue,citecolor=blue,allcolors=blue]{hyperref}
\usepackage{epsfig}
\usepackage{latexsym}
\usepackage[utf8]{inputenc}
\usepackage{xspace}
\usepackage{indentfirst}
\usepackage{enumitem}
\usepackage{color}

\usepackage{setspace}
\usepackage{lipsum}

\usepackage{amsmath}
\usepackage{amssymb}
\usepackage[english]{babel}
\usepackage{url}
\newcommand{\mean}[1]{\langle #1 \rangle}
\newcommand{\sk}[0]{S\sigma}
\newcommand{\kurt}[0]{\kappa\sigma^2}

\newcommand{\der}[2]{\frac{\partial #1}{\partial #2}}

\newcommand{\rom}[1]{\uppercase\expandafter{\romannumeral #1\relax}}

\newcommand{\eq}[1]{\begin{align} #1 \end{align}}

\begin{document}

\title{Critical point fluctuations: \\  Finite size and global charge conservation effects
}

\author{Roman V. Poberezhnyuk}
\affiliation{Bogolyubov Institute for Theoretical Physics, 03680 Kiev, Ukraine}
\affiliation{Frankfurt Institute for Advanced Studies, Giersch Science Center,
D-60438 Frankfurt am Main, Germany}

\author{Oleh Savchuk}
\affiliation{Physics Department, Taras Shevchenko National University of Kiev, 03022 Kiev, Ukraine}

\author{Mark I. Gorenstein}
\affiliation{Bogolyubov Institute for Theoretical Physics, 03680 Kiev, Ukraine}
\affiliation{Frankfurt Institute for Advanced Studies, Giersch Science Center, D-60438 Frankfurt am Main, Germany}

\author{Volodymyr~Vovchenko}
\affiliation{Nuclear Science Division, Lawrence Berkeley National Laboratory, 1 Cyclotron Road, Berkeley, California 94720, USA}
\affiliation{Frankfurt Institute for Advanced Studies, Giersch Science Center,
D-60438 Frankfurt am Main, Germany}

\author{Kirill~Taradiy}
\affiliation{Frankfurt Institute for Advanced Studies, Giersch Science Center,
D-60438 Frankfurt am Main, Germany}

\author{Viktor V. Begun}
\affiliation{University of Applied Sciences, An der Hochschule 1, 86161 Augsburg, Germany}
\affiliation{Warsaw University of Technology, Faculty of Physics, Koszykowa 75, 00-662 Warsaw, Poland}

\author{Leonid Satarov}
\affiliation{Frankfurt Institute for Advanced Studies, Giersch Science Center,
D-60438 Frankfurt am Main, Germany}

\author{Jan Steinheimer}
\affiliation{Frankfurt Institute for Advanced Studies, Giersch Science Center,
D-60438 Frankfurt am Main, Germany}

\author{Horst Stoecker}
\affiliation{Frankfurt Institute for Advanced Studies, Giersch Science Center, D-60438 Frankfurt am Main, Germany}
\affiliation{Institut f\"{u}r Theoretische Physik, Goethe Universit\"{a}t Frankfurt, D-60438 Frankfurt am Main, Germany}
\affiliation{
GSI Helmholtzzentrum f\"ur Schwerionenforschung GmbH, D-64291 Darmstadt, Germany}

\begin{abstract}

We investigate simultaneous effects of finite system size 
and global charge conservation on thermal fluctuations in the vicinity of a critical point. 
For that we consider a
finite interacting system which exchanges particles with a finite reservoir (thermostat),
comprising a statistical ensemble that is distinct from the common canonical and grand canonical ensembles.
As a particular example the van der Waals model is used.
The global charge conservation effects strongly influence the cumulants of particle number distribution when the system size is comparable to that of the reservoir.
If the system size is large enough to capture all the physics associated with the interactions,
the global charge conservation effects can be accurately described and corrected for analytically, within a recently developed subensemble acceptance method.
The finite size effects start to play a significant role when the correlation length grows large due to proximity of the critical point or when the system is small enough to be comparable to an eigenvolume of an individual particle.
We discuss our results in the context of fluctuation measurements in heavy-ion collisions.

\end{abstract}

\pacs{ }


\date{\today}

\maketitle

\section{Introduction}
The structure of the phase diagram 
of QCD matter is one of the most interesting unsolved problems in physics. 
Within  phenomenological statistical models as well as in lattice QCD simulations mainly the grand canonical ensemble (GCE) is used.  The critical behavior is then probed by the statistical fluctuations of  conserved charges 
\cite{Stephanov:1998dy,Stephanov:1999zu,Athanasiou:2010kw,Stephanov:2008qz,Kitazawa:2012at,Vovchenko:2015uda}.
Useful 
measures of these fluctuations are the scaled variance $\omega$, as well as the (normalized) skewness $S\sigma$ and kurtosis $\kappa\sigma^{2}$. For example, for the net baryon number $B$ they are defined as the following,
\eq{ \omega & =\frac{\langle (\Delta B)^2\rangle }{\mean{B}}~,~~~~
S\sigma  = \frac{\langle(\Delta B)^{3}\rangle}{\langle (\Delta B)^2 \rangle}~, \label{omegaB} \\
 \kappa  \sigma^{2} & = \frac{\langle(\Delta B)^{4}\rangle-3\langle(\Delta B)^{2}\rangle^2}{\langle (\Delta B)^2\rangle }~.\label{kurBt}
 }
 where $\langle...\rangle$ denotes the GCE  averaging and $ \Delta B \equiv  B-\langle B\rangle$. 
These quantities can also be expressed through baryon number cumulants $\kappa_n$:
\eq{\label{cumB}
\mean{B}=\kappa_1,~~~ \omega =\frac{\kappa_2}{\kappa_1},~~~  S\sigma=\frac{\kappa_3}{\kappa_2},~~~ \kappa\sigma^{2}=\frac{\kappa_4}{\kappa_2}.
}
The GCE  cumulants 
are calculated as the partial derivatives of the system pressure $p$  with respect to a corresponding chemical potential $\mu$:
\eq{\label{pres}
\kappa_n =VT^3~\frac{\partial^n (p/T^4)}{\partial (\mu/T)^n}~.
}
Here $V$ and $T$ are  the system volume and temperature, respectively.
The ratios of cumulants in Eq.~(\ref{cumB})
are intensive~(size-independent) measures in the GCE.

The GCE cumulants evaluated in effective QCD models can be directly compared with lattice QCD predictions, a procedure often used for testing and constraining various models and approaches~\cite{Borsanyi:2011sw,Bazavov:2012jq,Bhattacharyya:2013oya,Haque:2014rua,Fu:2016tey,Vovchenko:2016rkn,Critelli:2017oub,Vovchenko:2017gkg,Alba:2017bbr}.
On the other hand, a comparison of theoretical predictions with the event-by-event fluctuation measurements in relativistic heavy-ion collisions looks rather challenging. 
In the GCE the system of volume $V$ may exchange particles (and conserved charges) with a reservoir (thermostat) of volume $V_0-V$. In the total volume $V_0$ the conserved charge is strictly fixed. 
Thus, volume $V_0$ corresponds to a canonical ensemble (CE).
To reach the GCE conditions inside the volume $V$ one has   
to require $V/V_0  \ll 1$. 
And while direct comparisons of the GCE cumulants with experimental data are commonplace in the literature~\cite{Alba:2014eba,Fu:2016tey,Isserstedt:2019pgx,Bazavov:2020bjn},
it is clear that the global charge conservation will influence to some extent the conserved charge distribution measured in experiment, making it different from the GCE baseline.
Studies based on the ideal hadron gas model indeed show that higher-order cumulants of baryon number are strongly affected by the global conservation~\cite{Bzdak:2012ab}.
In addition, the volume $V$ should also be large enough to take into account all relevant physical effects due to particle interactions. 
If both $V_0$ and $V$ are large enough, one can derive analytically modifications, which come from global conservation laws. This has been shown in a recent paper \cite{Vovchenko:2020tsr} and will be discussed later in the present study. 

In high energy nucleus-nucleus collision experiments not all final particles are measured
on an event-by-event basis. 
Within a statistical approach the subset of measured particles can be treated as a subsystem with finite volume $V$, whereas nondetected particles play the role of the finite reservoir (thermostat).
In this situation, the effects of exact charge conservation on fluctuations are usually modeled by a binomial acceptance correction 
procedure \cite{Bzdak:2012ab,Kitazawa:2011wh,Kitazawa:2016awu,Braun-Munzinger:2016yjz,Savchuk:2019xfg}. This procedure  assumes that the probability to be measured is the same for each particle of a given type and it is independent of any inter-particle correlations.
As will be seen below, the binomial acceptance  procedure can be justified only for a statistical system of classical non-interacting particles.

Previously, the finite size effects (without conservation law effects) for the first order liquid-gas phase transition were discussed in Refs.~\cite{CSERNAI199425,Spieles:1997ab,Bzdak:2018uhv,Spieles:2019ynp}. 
The effects of finite particle number sampling on baryon number fluctuations 
were studied in Ref.~\cite{Steinheimer:2017dpb} within fluid dynamical simulations.
A Monte Carlo procedure allowing to sample particle multiplicities 
in the presence of excluded volume effects was developed in Ref.~\cite{Vovchenko:2018cnf}. 

The size of the considered system becomes 
especially important in the vicinity of the critical point (CP).  
The CP as the end point of the first-order phase transition exists as a universal feature of all molecular systems. 
At the CP the intensive fluctuation measures  become singular in the thermodynamic limit $V\rightarrow \infty$. 
These infinite values evidently cannot appear in a finite $V$. 
Therefore, both the charge conservation and finite size effects
can be equally important in the vicinity of the CP.

It was demonstrated \cite{Vovchenko:2016rkn} that the nuclear CP, i.e., the end point of the liquid-gas transition in the system of interacting nucleons at small $T$ and large $\mu_B$, affects the susceptibilities of conserved charges even at $\mu_B=0$ and large $T$, and limits the radius of convergence of Taylor expansion in $\mu_B/T$ at $\mu_B=0$~\cite{Savchuk:2019yxl}.
The sought-after hypothetical chiral QCD CP is expected to produce strong signals in high-order fluctuation measures~\cite{Stephanov:1999zu,Hatta:2002sj,Stephanov:2008qz,Stephanov:2011pb}.
It is possible that conserved charge susceptibilities are determined by a complex interplay of the chiral and liquid-gas phase transitions in certain regions of the phase diagram~\cite{Mukherjee:2016nhb,Motornenko:2019arp}.

Our paper presents a first step to study both the finite size and global charge conservation effects in the vicinity of a CP.
To give a specific example we consider a classical statistical system of interacting nucleons described by the van der Waals (vdW) model. 
This model  was previously applied to nuclear matter considered as  a system of interacting nucleons in Ref.~\cite{Vovchenko:2015vxa}.
The production of antibaryons will be neglected. 
In this situation the number of nucleons becomes a conserved charge.
We will not consider the mixed phase region at $T<T_c$ in the present study, and will focus our studies at (super)critical temperatures, $T \geq T_c$.

Our considerations will be based purely on equilibrium statistical mechanics.
The non-equilibrium effects in heavy-ion collisions are certainly important, especially in the vicinity of the CP, and the dynamical theory of critical fluctuations is under development~\cite{Mukherjee:2015swa,Stephanov:2017ghc,Nahrgang:2018afz,Akamatsu:2018vjr,Bluhm:2020mpc}. We plan to incorporate the non-equilibrium effects in future works.

The paper is organized as follows. 
Section~\ref{sec-vdW} presents the main properties of the vdW model in the thermodynamic limit.   
Section~\ref{model} describes the model results for the particle number fluctuations in the finite systems.
The summary in Sec.~\ref{summary} closes the paper.

\section{van der Waals model}
\label{sec-vdW}

The CE partition function, $Z_{\rm ce}$, for the 
vdW system of classical   particles 
can be written as~\cite{GNS}  
\eq{\label{zce}
Z_{\rm ce}(N,V,T)& =\frac{[\varphi(T)]^{N}}{N!}\,(V-bN)^{N}\\
& \times~ \theta(V-bN)\, \exp\left(\frac{aN^2}{VT}\right)~, \nonumber
}
where $N$, $V$, and $T$ are, respectively, the number of particles,  volume, and the temperature of the system, while $a>0$ and $b>0$ are the vdW interaction parameters.  The $a$ parameter regulates the attraction, while $b$ corresponds to a repulsion between particles via the excluded volume effects. 
The function $\varphi$ is given as
\eq{
\varphi(T;g,m)\equiv \frac{g}{\pi^2}\,T\,m^2\,K_2(m/T)~,
}
where $g$ and $m$ are, respectively, the degeneracy factor and the mass of the particles, and $K_2$ is the modified Bessel function of the second kind.

The system pressure in the CE is calculated as
\eq{\label{pressure}
p(N,V,T) =T\left(\frac{\partial \ln Z}{\partial V}\right)_{N,T}= \frac{n\,T}{1-b\,n}-an^2~ ,
}
where $n\equiv N/V$.
The CP is defined by the conditions~\cite{GNS,LL} 
\eq{\label{cp}
\left(\frac{\partial p}{\partial n}\right)_T=0~,~~~~\left(
\frac{\partial^2 p}{\partial n^2}\right)_T=0~,
}
which gives  
\eq{\label{cp-1}
T_c= \frac{8a}{27 b}~,~~~~ n_c=\frac{1}{3b}~,~~~~p_c=\frac{a}{27b^2}~.
}
Introducing the reduced variables $\widetilde{T}=T/T_c$, $\widetilde{n}=n/n_c$, and  $\widetilde{p}=p/p_c$ one can rewrite the vdW equation (\ref{pressure}) in a universal form
\eq{\label{vdW-p}
\left(\widetilde{p}~+~3\,\widetilde{n}^2 \right)\,\left(\frac{3}{\widetilde{n}}~-~1\right)=~\widetilde{T}~,
}
which is independent of the specific numerical values of the interaction parameters $a$ and $b$.
This is a particular case of the principle of the corresponding states (see, e.g., Ref.~\cite{GNS}).

To calculate the particle number fluctuation measures one usually transforms  the CE description into the GCE one. This requires to introduce a reservoir and to take the thermodynamic limit with $ V\rightarrow \infty$.
For the vdW equation of state these steps  were  done for the first time  in Ref.~\cite{Vovchenko:2015xja}. In the vdW model the particle number fluctuation measures  
\eq{ 
\omega & =\frac{\langle (\Delta N)^2\rangle}{\langle N\rangle}~,
\label{omega}
~~~~~
S\sigma  = \frac{\langle(\Delta N)^{3}\rangle}{\langle (\Delta N)^2 \rangle}~,\\
 \kappa  \sigma^{2} & = \frac{\langle(\Delta N)^{4}\rangle-3\langle(\Delta N)^{2}\rangle^2}{\langle (\Delta N)^2\rangle} ~,\label{kurt}
 }
with $\Delta N\equiv N-\langle N\rangle$,
were calculated analytically in the GCE in the thermodynamic limit $V\rightarrow \infty$~\cite{Vovchenko:2015xja,Vovchenko:2015uda}:
\eq{
\label{omega-gce}
\omega_{\rm gce} 
&~=~\frac{1}{9}\left[\frac{1}{(3-\widetilde{n})^2}-\frac{\widetilde{n}}
{4 \widetilde{T}}\right]^{-1}~, \\
\label{S-gce}
S\sigma_{\rm gce} &~=~ 
\frac{1}{3}\left[\frac{1}{(3-\widetilde{n})^2}-\frac{\widetilde{n}}{4 \widetilde{T}}\right]^{-2}~
\left[\frac{1-\widetilde{n}}{(3-\widetilde{n})^3}\right]~,
\\
\label{kurt-gce}
\kappa \sigma^2_{\rm gce} &~=~
3\, (S\sigma)^2 - 2\, \omega \, S\sigma-
54\, \omega^3\frac{\widetilde{n}^2}{(3-\widetilde{n})^4}~.
}
\begin{figure}[h!]
\includegraphics[width=.49\textwidth]{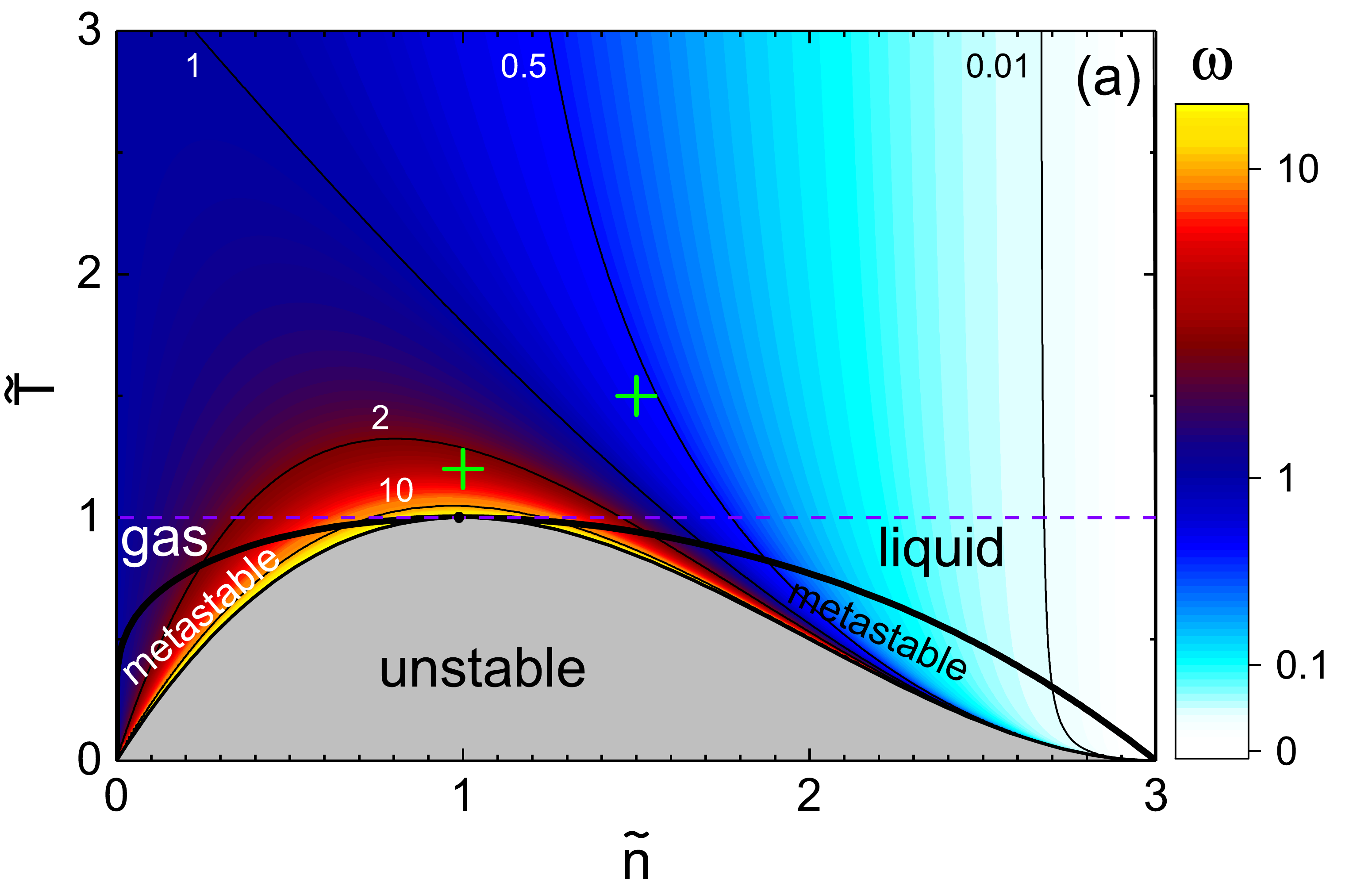}
\includegraphics[width=.49\textwidth]{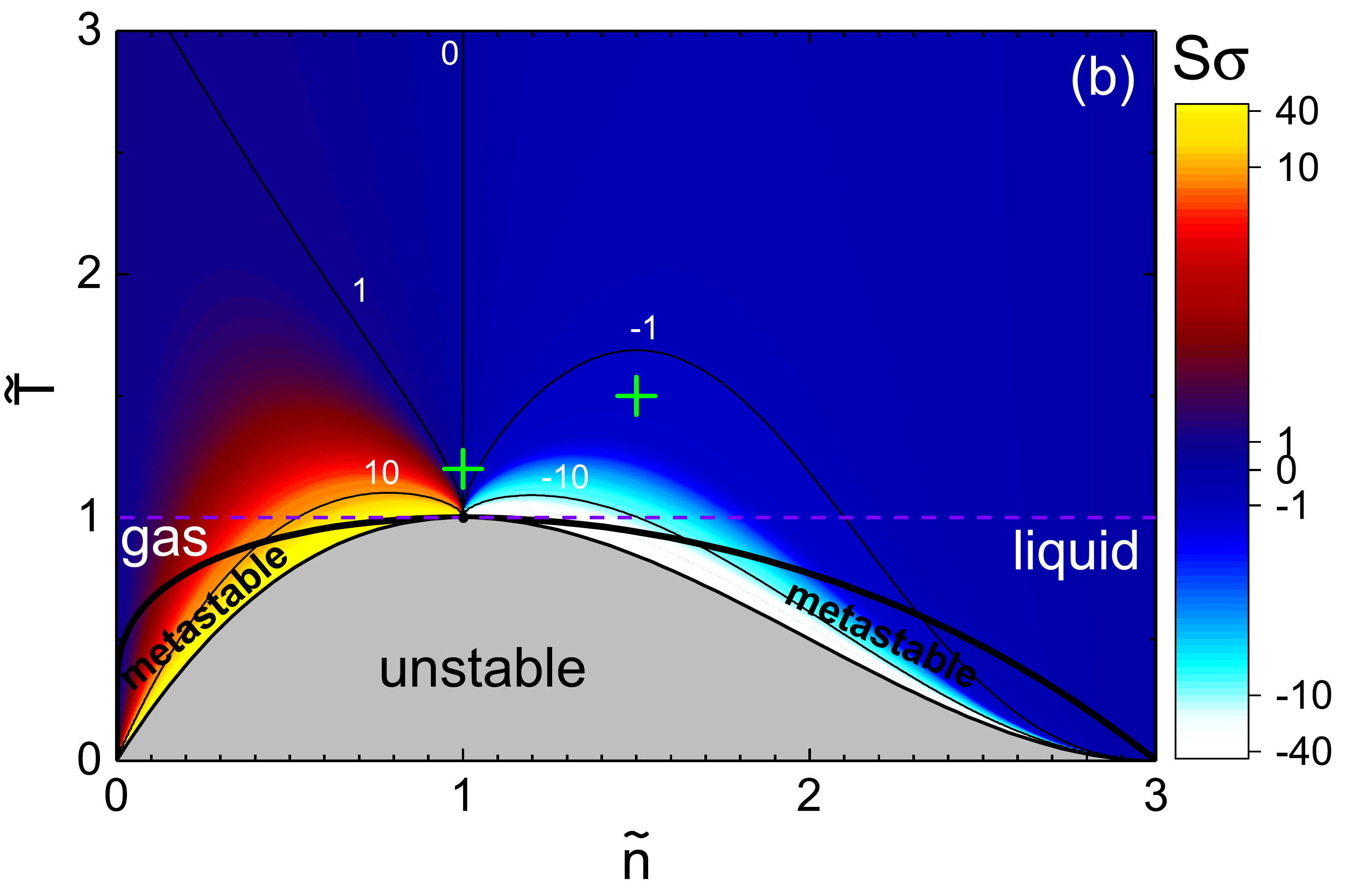}
\includegraphics[width=.49\textwidth]{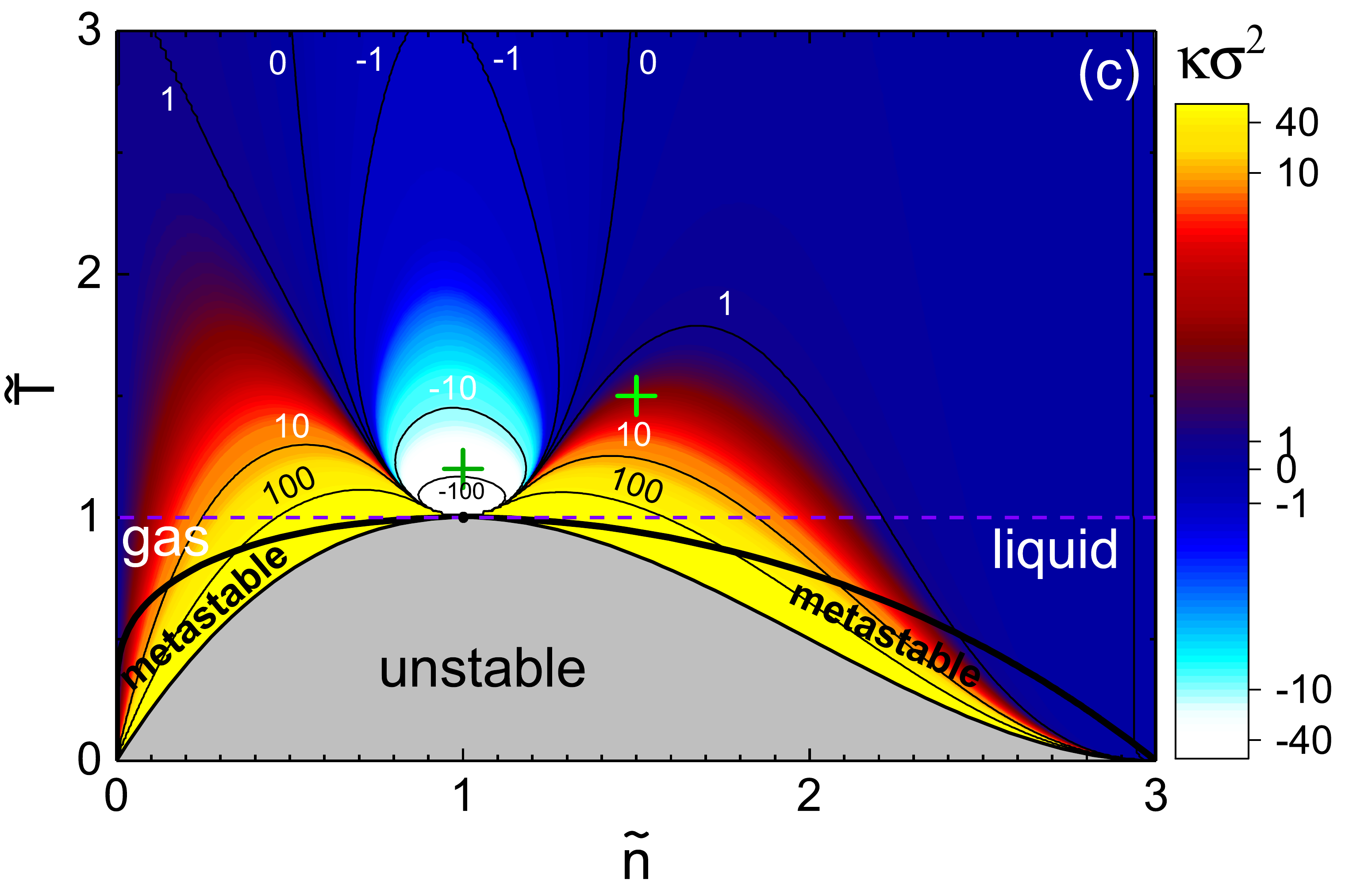}
\caption{\label{fig-gce-fluct}
The GCE fluctuation measures  ($a$) $\omega_{\rm gce}$, ($b$) $S\sigma_{\rm gce}$, and ($c$) $\kurt_{\rm gce}$  given by Eqs.~(\ref{omega-gce})--(\ref{kurt-gce})
in the $(\widetilde{n},\widetilde{T})$  plane. 
The Poisson limit with $\omega_{\rm gce}=S\sigma_{\rm gce}=\kurt_{\rm gce}=1$
corresponds to those regions of the thermodynamic plane where the vdW interactions become negligible. The two points on the phase diagram marked by crosses are analyzed in detail in the present paper.
}
 \end{figure}
The GCE fluctuation measures \eqref{omega-gce}-\eqref{kurt-gce} are presented in Fig.~\ref{fig-gce-fluct}.
All three 
of them exhibit singular behavior at the CP.
While the $\omega_{\rm gce}$ (\ref{omega-gce}) tends to  $+\infty$ at the CP, the $\sk_{\rm gce}$ (\ref{S-gce}) and $\kurt_{\rm gce}$ (\ref{kurt-gce}) have a richer structure in a vicinity of the CP.
They can tend to $+\infty$, $-\infty$, or $0$ depending on the path of the approach to the CP.
Introducing quantities $\rho=\widetilde{n}-1$ and
$\tau=\widetilde{T}-1$ one finds at $\tau\ll 1$ and $\rho\ll 1$:
\eq{
&\omega_{\rm gce}\cong\frac{4}{9}\left[\tau+\frac{3}{4}\rho^2+\tau\rho\right]^{-1}~, \label{omega-gce-1}\\
&S\sigma_{\rm gce}\cong -~\frac{2}{3}\rho\,\left[\tau+\frac{3}{4}\rho^2+\tau\rho\right]^{-2}~,\label{S-gce1} \\
&\kappa\sigma^2_{\rm gce}\propto\rho^{-6}~{\rm at}~\tau=0,~~~~\kappa \sigma^2_{\rm gce}\propto-\tau^{-3}~{\rm at}~\rho=0~.
\label{kurt-gce1}
}

While the CP signals of $\omega$ fade out as one moves away from the CP in the phase diagram, they remain stronger in the higher-order fluctuation measures, $\sk$ and $\kurt$, even far away from the CP \cite{Poberezhnyuk:2019pxs}.
Note that in the classical ideal gas case, $a=b=0$, all fluctuation measures in Eqs.~(\ref{omega-gce})--(\ref{kurt-gce}) are reduced to   $\omega_{\rm gce}=\sk_{\rm gce}=\kurt_{\rm gce}=1$, which corresponds to the Poisson  $N$-distribution.
The general features of the GCE fluctuations presented in this section, especially those connected with the CP  remain the same for all models from the mean-field universality class, to which the vdW model belongs~(see, e.g., Ref.~\cite{Poberezhnyuk:2017yhx}).

\section{Fluctuations in a subensemble}
\label{model}
Let us partition a finite system of volume $V_0$ into
 a subsystem of volume $V<V_0$ and another subsystem -- a reservoir -- of volume $V_0-V$. We assume  that both subsystems can 
exchange particles, but the total number of particles $N_0$ in the whole system is fixed. 
The corresponding statistical  ensemble will be referred as a {\it subensemble}, distinguishing it from both the CE and GCE.
We neglect all interactions at the interface, i.e. between all particles from different subsystems.
The partition function of the system in volume $V$ can then be written as~\cite{GNS,Vovchenko:2020tsr}:  
\eq{\label{partition2}
Z(V,T)  & =\sum_{N=N_{\rm min}}^{N_{\rm max}} Z_{\rm ce}(V,N,T) Z_{\rm ce}\left(V_0-V,N_0-N,T\right)~  \nonumber \\
 &\equiv \sum_{N=N_{\rm min}}^{N_{\rm max}}Z(N;V,T)~.
}
The probability to find $N$ particles in the volume 
$V$ takes the form
\eq{\label{probability}
W(N;V,T)~=~\frac{Z(N;V,T)}{Z(V,T)}~.
}
The mean value $\langle N\rangle$ and the central moments 
$\langle (\Delta N)^k\rangle $ with $k=2,3,\ldots$ in the subensemble are calculated
with the probability distribution (\ref{probability}):
\eq{\label{NkW}
\langle \ldots \rangle=\sum_{N=N_{min}}^{N_{max}}\ldots \,W(N;V,T)
}

In our example of the vdW model, the CE partition functions 
$Z_{\rm ce}$ in Eq.~(\ref{partition2}) are given by Eq.~(\ref{zce}). Introducing variables 
\eq{
n\equiv N_0/V_0~~~~~ {\rm and}~~~~~  x\equiv V/V_0 
}
the partition function~(\ref{partition2}) is written as
\eq{
Z(V,T)  & \equiv \sum_{N=N_{\rm min}}^{N_{\rm max}}Z(N;V,T) \nonumber \\
& = \eta^{N_0}\sum_{N=N_{\rm min}}^{N_{\rm max}}\frac{1}{N!(N_0-N)!}\left(\frac{3}{\widetilde{n}}x-\frac{N}{N_0}\right)^N
\nonumber \\
& \quad \times \left[\frac{3}{\widetilde{n}}(1-x)+\frac{N}{N_0}-1\right]^{N_0-N} \nonumber \\
& \quad  \times \exp\left[\frac{9}{4}\frac{\widetilde{n}}{\widetilde{T}}\frac{N}{1-x}\left(\frac{1}{2x}\frac{N}{N_0}-1\right)\right]~,
\label{z-subensemble}
}
Here
\eq{
\eta=b~N_0~\varphi(T;g,m)~ \exp\left[
\frac{9\,\widetilde{n}}{8\,\widetilde{T}\,(1-x)}\right],
}
and it cancels out in the  probability distribution~(\ref{probability}).

The minimal and maximal numbers of particles in the subensemble, $N_{\rm min}={\rm max}\{0,\,N_0-\lfloor(V_0-V)/b\rfloor\}$ and
 $N_{\rm max}={\rm min}\{\lfloor V/b \rfloor,\,N_0\}$ result from the Heaviside $\theta$-functions in $Z_{\rm ce}$ (\ref{zce}), which is due to the excluded volumes. 
 Here  $\lfloor...\rfloor$ is a floor function. $N_{\rm min}$ and $N_{\rm max}$ can also be rewritten as:
\eq{
&N_{\rm min}={\rm max}\{0,\,N_0-\lfloor 3(1-x) N_0 /\widetilde{n}\rfloor\},\\
&N_{\rm max}={\rm min}\{\lfloor 3x N_0 /\widetilde{n}\rfloor,\,N_0\}.
}

The  moments  (\ref{NkW})  are independent of particles degeneracy $g$ and mass $m$, since they only enter Eq.~\eqref{z-subensemble} through the common factor $\eta$\footnote{This would not be the case if the quantum statistics was not neglected~\cite{Vovchenko:2015xja}.}.
In the following we explore the behavior of fluctuations in the subensemble for different values of $x$ and $N_0$.

\subsection{Charge conservation effects }
As a first specific case, we consider the   thermodynamic limit, $V_0\rightarrow\infty$, at $0<x<1$. 
Thus, both $V_0\rightarrow \infty$ and $V\rightarrow \infty$, but the values of $x=V/V_0$ remain finite. 

In this case we follow a recently developed subensemble acceptance procedure~\cite{Vovchenko:2020tsr}. 
It allows to obtain the cumulants of particle number distribution in the subensemble 
in terms of the corresponding
GCE cumulants and the volume fraction $x$, 
quantifying the corrections to the GCE cumulants because of the global conservation of particle number.
One obtains~(see Ref.~\cite{Vovchenko:2020tsr} for the derivation details):

\begin{subequations}
\label{cum-gce}
\begin{align}
\kappa_1 &= \mean{N}=\xi_1 N_0,
\\
\kappa_2 & = \xi_2\kappa^{\rm gce}_{2},
\\
 \kappa_3 & = \xi_3 \kappa^{\rm gce}_{3}, 
 \\
\kappa_4 & = \xi_4\kappa^{\rm gce}_4 + 3\xi_2^2\frac{\kappa^{\rm gce}_4 \kappa^{\rm gce}_2-[\kappa^{\rm gce}_3]^2}{\kappa^{\rm gce}_2}~,
\end{align}
\end{subequations}
where $\kappa^{\rm gce}_n\propto V_0$ is the $n$-th cumulant in the GCE, and 
\begin{subequations}
\label{cum-binomial}
\begin{align}
\xi_1& =x, \\ 
\xi_2& =x (1-x),\\ 
\xi_3& =\xi_2 (1-2x), \\ 
\xi_4& =\xi_2(1-6\xi_2)~. 
\end{align}
\end{subequations}
Note that $\xi_n(x)$ correspond to the $n$-th cumulant of the Bernoulli distribution, $p_x(l) = x^l(1-x)^{1-l}$ for $l=0,\,1$.
Similarly, higher order $\kappa_n$ and $ \xi_n$ cumulants can be obtained.
Using Eqs.~(\ref{cum-gce}) and (\ref{cum-binomial}) one finds the scaled variance, skewness, and kurtosis:
\eq{\label{omega-tdl}
\omega & =(1-x)\,\omega_{\rm gce}~,\\
\sk   &  =(1-2 x)\,\sk_{\rm gce}~,\\
\label{kurt-tdl}
\kurt &  =(1-6 \xi_2)\,\kurt_{\rm gce}+3\xi_2\,[\kurt_{\rm gce}-(\sk_{\rm gce})^2]~.
}
Equations (\ref{omega-tdl})-(\ref{kurt-tdl}) present the intensive measures of particle number fluctuations in the subensemble
in terms of the corresponding GCE cumulant ratios. This  greatly simplifies   the consideration as the finite size effects are neglected. At finite $x$-values the fluctuation measures (\ref{omega-tdl}) and (\ref{kurt-tdl}) are still influenced by the global conservation of $N_0$. These global conservation effects, however, are expressed as universal functions of $x$.
The expressions (\ref{omega-tdl}) and (\ref{kurt-tdl})  are model independent \cite{Vovchenko:2020tsr}. 
Moreover, they are valid not only for particle number fluctuations but also for fluctuations of a conserved charge, e.g., for the fluctuation measures of the net baryon charge in Eqs.~(\ref{omegaB})-(\ref{kurBt}). 

The skewness and kurtosis  are, respectively, anti-symmetric and symmetric functions around $x=1/2$, i.e.,
$\sk[1-x]=-\sk[x]$ and $\kurt[1-x]=\kurt[x]$.
The fluctuation measures (\ref{omega-tdl})--(\ref{kurt-tdl}) reduce to the GCE ones~ (\ref{omega-gce})--(\ref{kurt-gce}) in the limit $x\rightarrow 0$\footnote{Note that we still assume here that the volume $V$ is large enough to neglect the finite-size effects, no matter how small $x$ is.}. 
On the other hand, at $x\rightarrow 1$ the cumulant ratios approach   
\eq{
\omega & \stackrel{x \to 1}{=} 0~,\\
\sk &  \stackrel{x \to 1}{=}~-\sk_{\rm gce}~,\\
\kurt &  \stackrel{x \to 1}{=}\kurt_{\rm gce}~.
}

\begin{figure}
\includegraphics[width=.49\textwidth]{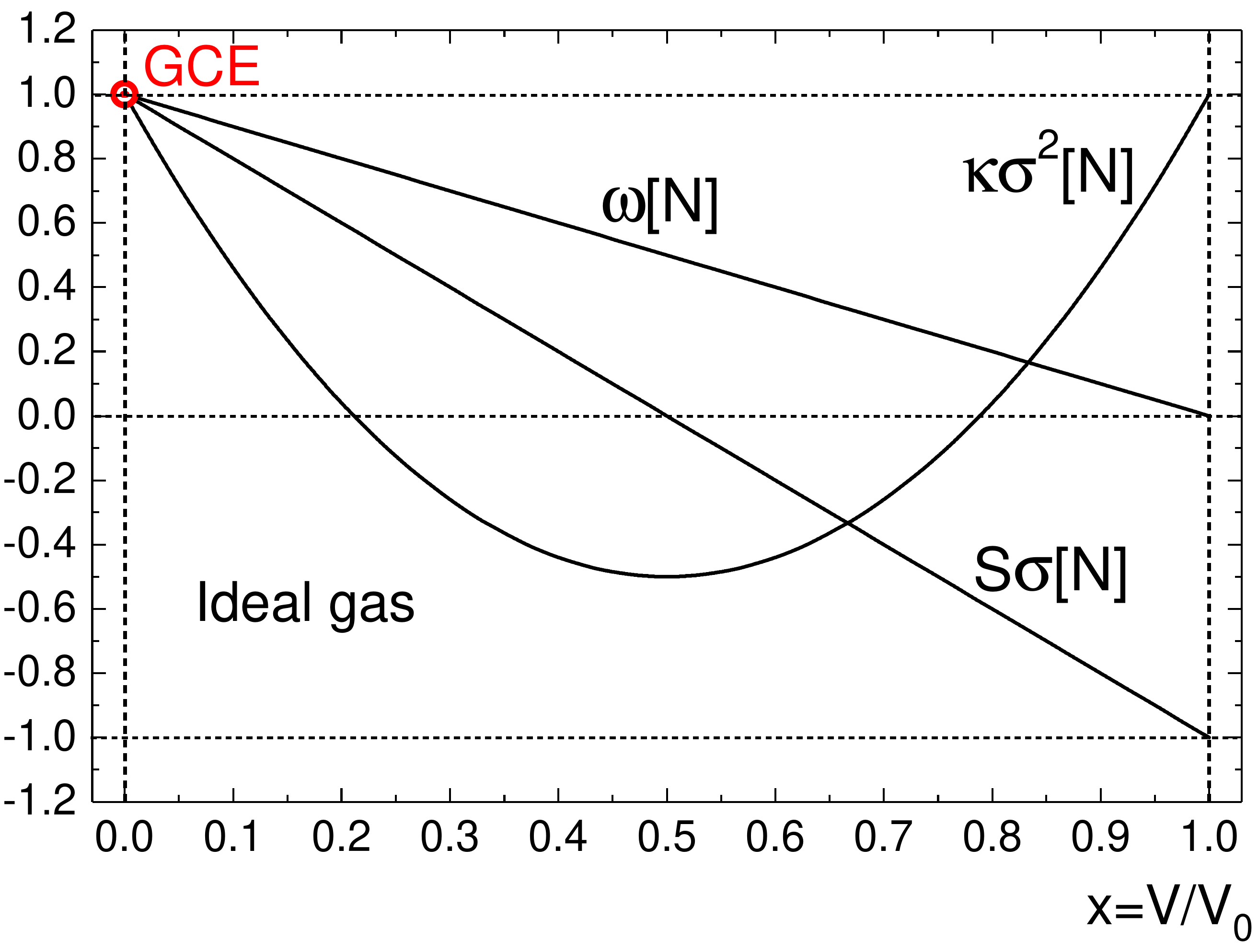}
\caption{\label{fig-ideal}
Scaled variance $\omega_{\rm id}$, skewness $\sk_{\rm id}$, and kurtosis $\kurt_{\rm id}$ of particle number fluctuations for a classical ideal gas of particles in a subvolume as a function of a fraction $x$ of the total volume which is occupied by the subvolume. 
The grand-canonical ensemble values correspond to the red circle.
}
 \end{figure}

It is instructive to consider the limit of an ideal classical gas.
This limit is recovered for $a=0$ and $b=0$.
In this case, $\kappa^{\rm gce}_n=\mean{N_0}$ for all~$n=1,2,\ldots $, and Eqs.~(\ref{cum-gce}) reduce to
$\kappa^{\rm id}_n=\xi_n\mean{N_0}$. 
Therefore, one obtains the following for the classical ideal gas:
\eq{\label{ideal}
\omega_{\rm id} & = 1-x,\\
\sk_{\rm id} & = 1-2x,\\
\label{idealkurt}
\kurt_{\rm id} & = 1-6x(1-x).
}
The fluctuation measures \eqref{ideal}-\eqref{idealkurt} are presented in Fig.~\ref{fig-ideal}. 
Equations~\eqref{ideal}-\eqref{idealkurt} 
coincide with those obtained after the binomial acceptance correction procedure (see Ref.~\cite{Savchuk:2019xfg} for details). 
The binomial acceptance procedure is suitable for describing the global charge conservation effects in non-interacting systems.
The applicability of the binomial acceptance, however, does not extend to interacting systems, the vdW model in particular.

If the system is close to the thermodynamic limit, Eqs.~(\ref{omega-tdl})-(\ref{kurt-tdl}) can be used to account for global conservation effects. The requirements for the system to be close to the thermodynamic limit depend on the specific properties of the system under consideration. Previously, it was demonstrated that for the non-interacting Hadron Resonance Gas the scaled variance of particle number fluctuations is close to it's thermodynamic limit values already for $\langle N \rangle\approx 2-3$, see e.g. Ref.~\cite{Begun:2004gs}.
However, the size of the interacting system near the CP must be larger for the thermodynamic limit to be applicable. We will investigate the requirements for such a system in Sec.~\ref{sec-general} by comparing the thermodynamic limit results with the direct finite-size calculations within the vdW model for different system sizes.

 \begin{figure*}
\includegraphics[width=.49\textwidth]{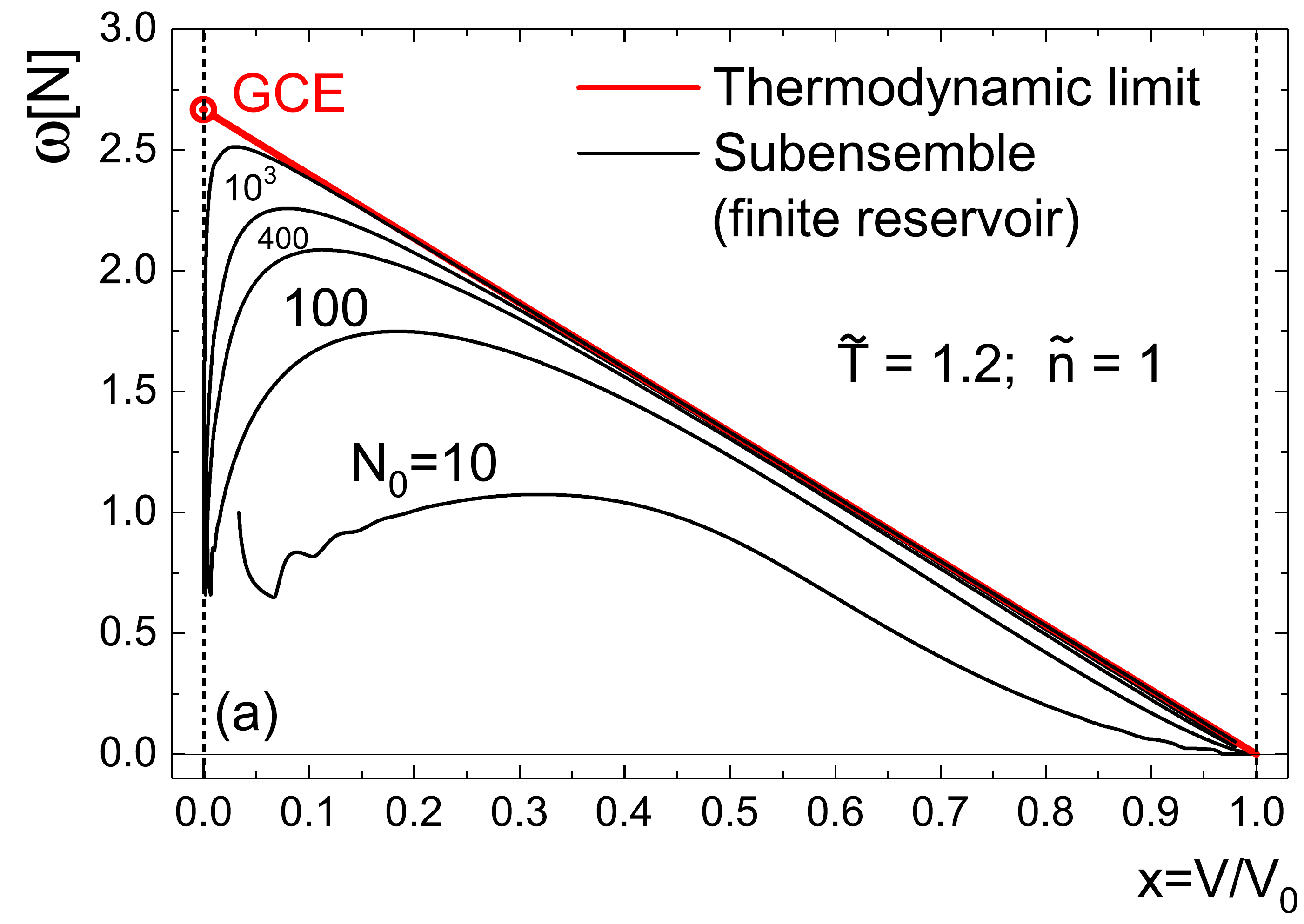}
\includegraphics[width=.49\textwidth]{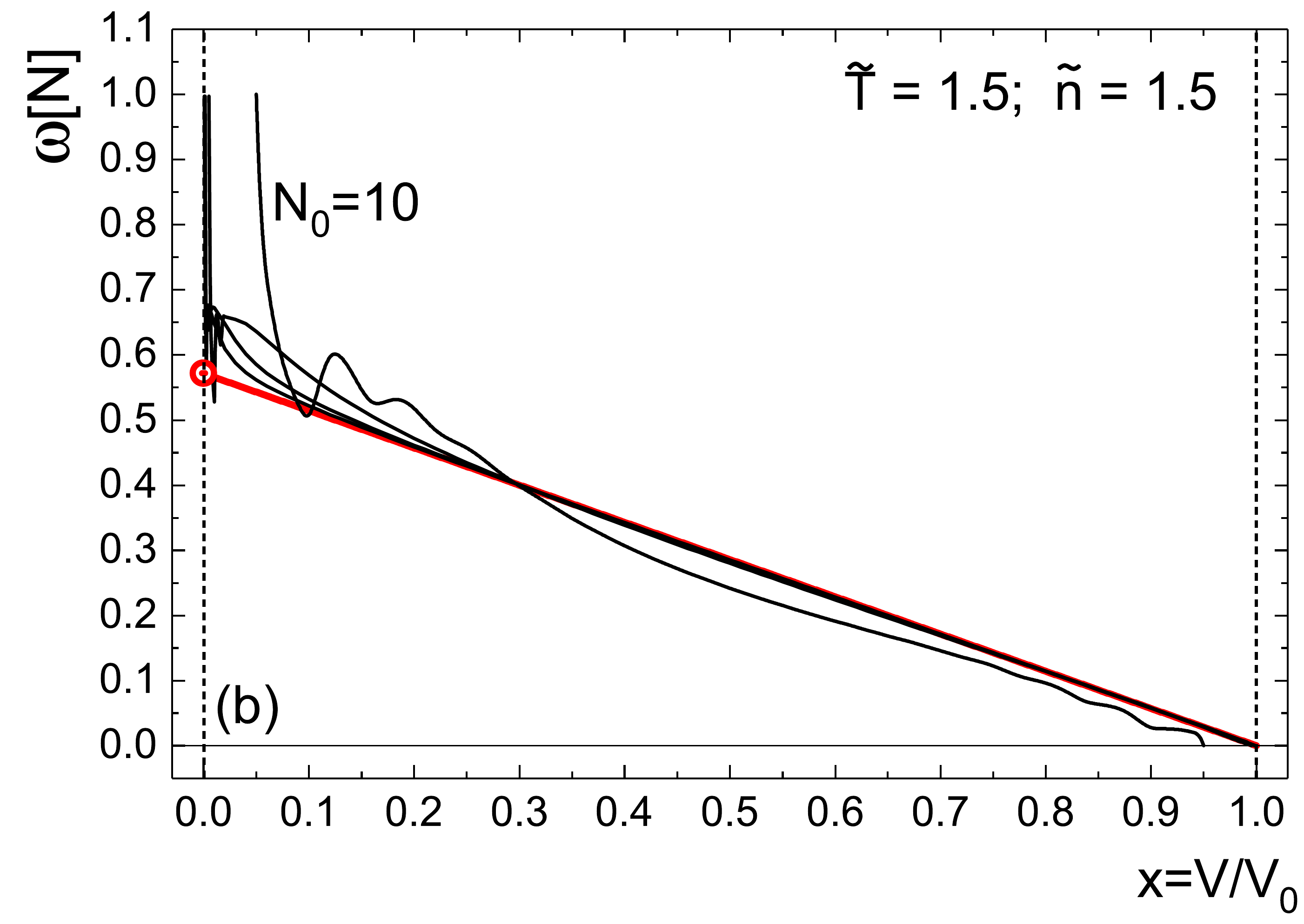}
\includegraphics[width=.49\textwidth]{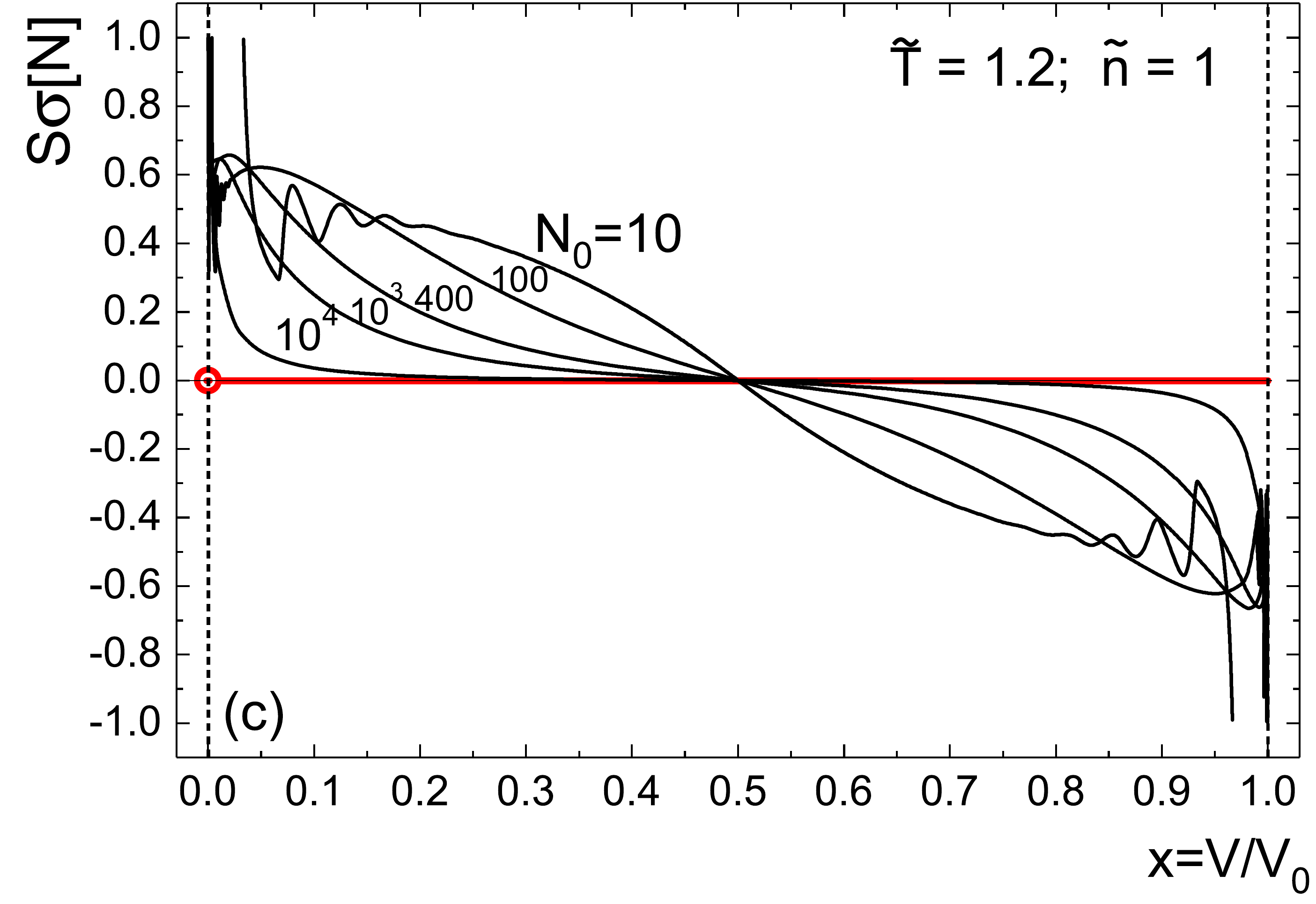}
\includegraphics[width=.49\textwidth]{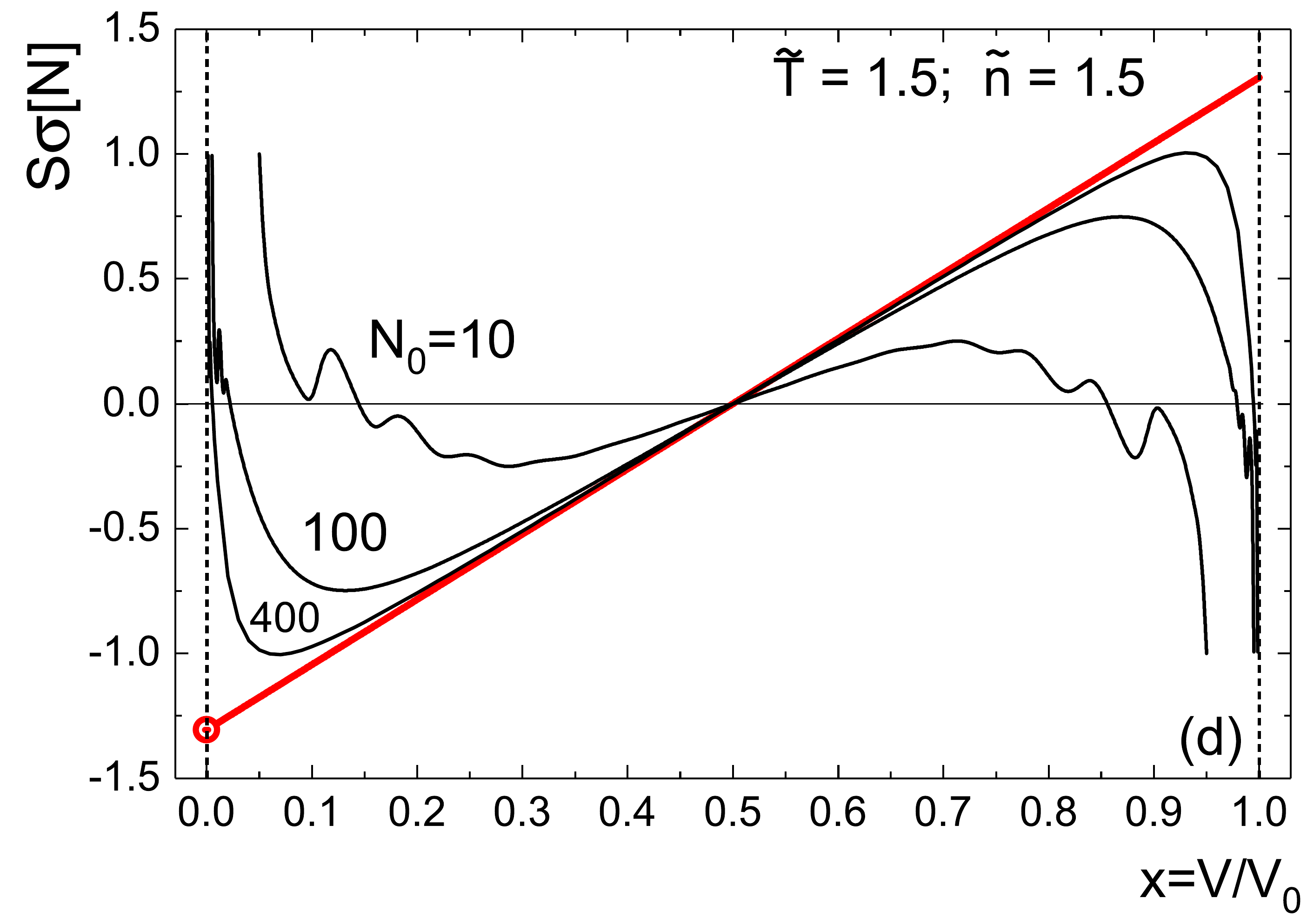}
\includegraphics[width=.49\textwidth]{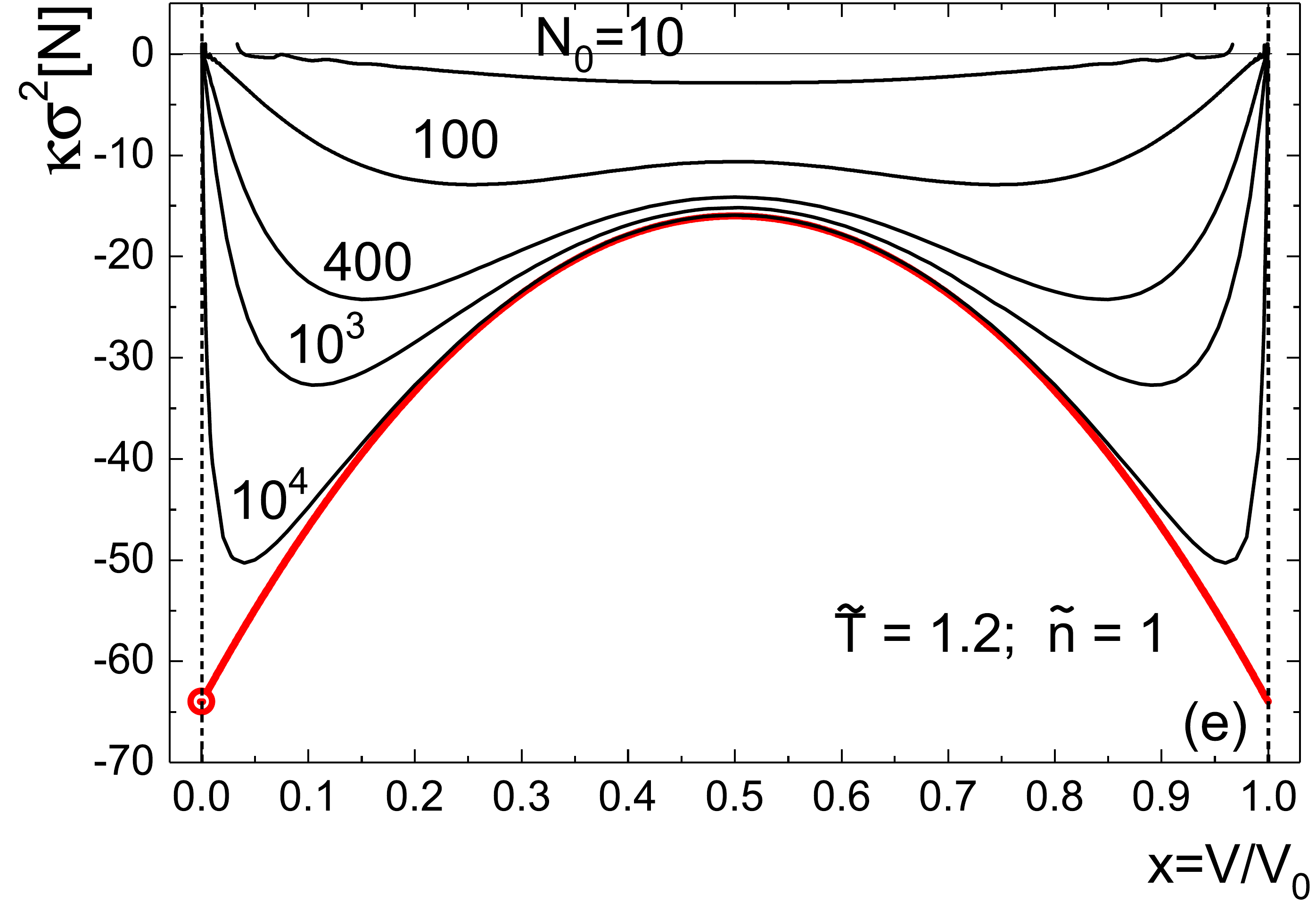}
\includegraphics[width=.49\textwidth]{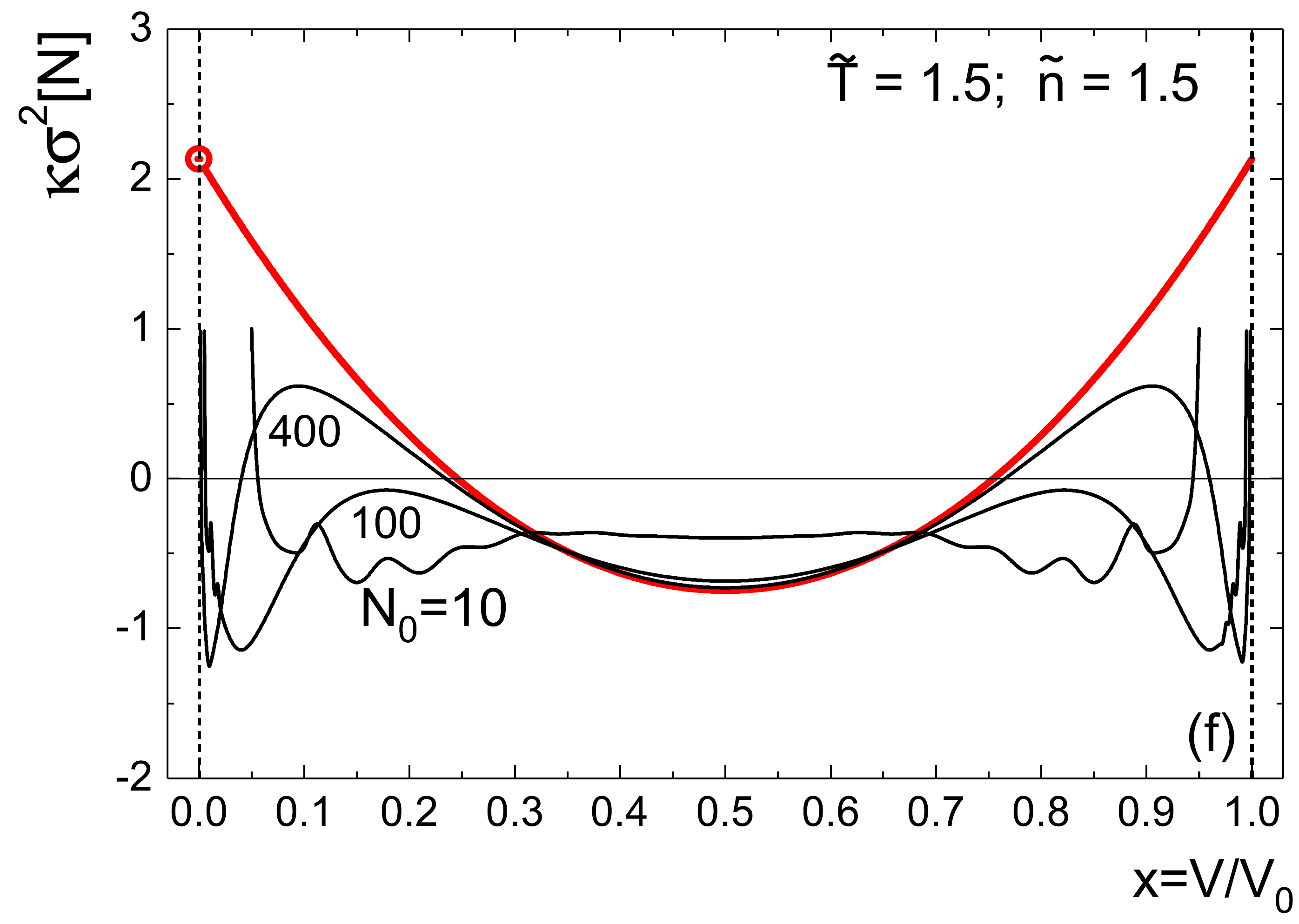}
\caption{\label{general}
(a),(b) Scaled variance $\omega$, 
(c),(d) skewness $\sk$, and (e),(f)
kurtosis $\kurt$ of particle number fluctuations in the subensemble with volume $V$ are presented as functions of $x$. 
Results for 
$(\widetilde{n}= 1, \widetilde{T}=1.2)$ and 
$(\widetilde{n}=1.5, \widetilde{T}=1.5)$ are shown in the left [(a),(c),(e)] and right [(b),(d),(f)] panel, respectively.  Different values of $N_0$
between $10$ and $10^4$ are presented. Red lines show the thermodynamic limit, $N_0\rightarrow\infty$, given by Eqs.~(\ref{omega-tdl}) and (\ref{kurt-tdl}) of the subensemble acceptance method of Ref.~\cite{Vovchenko:2020tsr}. 
The GCE results (\ref{omega-gce})-(\ref{kurt-gce}) are depicted by the red circles.
}
 \end{figure*} 

\subsection{General case of a finite reservoir }
\label{sec-general}

In this subsection we consider the general case when both the system and reservoir are finite. Thus, both the finite size and conservation law effects are present.
Figure~\ref{general} shows examples of the subensemble particle number fluctuations calculated 
according to the general Eqs.~(\ref{probability})--(\ref{z-subensemble})
at finite $N_0$. Different black lines  show different $N_0$ values
from
$N_0=10$ to $N_0=10^4$.
Two locations in the phase diagram, $(\widetilde{n}=1,\,\widetilde{T}=1.2)$ [Figs.~\ref{general}(a), \ref{general}(c), and \ref{general}(e)] and $(\widetilde{n}=1.5,\,\widetilde{T}=1.5)$ 
[Figs.~\ref{general}(b), \ref{general}(d), and \ref{general}(f)] are considered. These two points are marked by crosses in Fig.~\ref{fig-gce-fluct}. 
The choice of these specific points for the illustration is due to the following reasons. First, these two locations correspond to rather different GCE  values for the  fluctuation measures, which are shown by full red circles in Fig.~\ref{general}. 
The deviations from the ideal gas limit in both cases are large.
Second, these two points in the 
$\widetilde{n}$-$\widetilde{T}$ plane 
are in different proximities to the CP at $(\widetilde{n}=1,\,\widetilde{T}=1)$.

The thermodynamic limit results given by Eqs.~(\ref{omega-tdl})--(\ref{kurt-tdl}) are represented by red lines in Fig.~\ref{general}. The points  $x\rightarrow 0$ on these red lines correspond to the  GCE values \eqref{omega-gce}--\eqref{kurt-gce}. 
They are shown in Fig.~\ref{general} by  full red circles. 
The $x$-dependence  according to Eqs.~(\ref{omega-tdl})--(\ref{kurt-tdl}), shown by the red lines in Fig.~\ref{general}, reflects the global $N_0$ conservation. 
The comparison of these lines with those in Fig.~\ref{fig-ideal} shows a strong
sensitivity of the skewness and kurtosis to the presence of interactions between particles.
At finite $x$ values the effects of the $N_0$-conservation keep being significant even in the thermodynamic limit $N_0\rightarrow \infty$. At finite $N_0$ there are additional finite size effects.  How large $N_0$ should be to approach the thermodynamic limit shown by red lines in Fig.~\ref{general} with a certain accuracy? 
This depends on both the proximity of the point $(\widetilde{n},\widetilde{T})$ to the CP on the phase diagram and the numerical value of $x$. 
The closer the system is to the CP, the larger are the finite-size effects.
This evidently reflects the growth of the correlation length as one approaches the CP, which is known to become of a macroscopic magnitude at the CP.

The magnitude of the finite size effects at a fixed $N_0$ is minimal at $x=1/2$, as seen from  Fig.~\ref{general}. 
This is because both volumes $V=V_0/2$ and $V_0-V=V_0/2$ are relatively large in this case. 
Thus, to minimize the finite size effects in the  event-by-event fluctuation data  
it may be worthwhile to aim for an acceptance, which encompasses close to $50\%$ of all final particles on average.
The effects from the global charge conservation are not small in this case. However, they can be estimated (and then corrected for) using the formulas of the subensemble acceptance procedure,
Eqs.~(\ref{omega-tdl})--(\ref{kurt-tdl}).
It should be noted, however, that the skewness goes to zero at $x=1/2$, as this quantity  is an asymmetric function of $x$ in the interval $[0,1]$.
It would therefore be necessary to consider acceptance away from $x = 1/2$ for this quantity.

The thermodynamic limit can be reached also at smaller $x$-values.
The smaller $x$-values would, however, require the larger $N_0$ to reach the same level of accuracy with respect to the finite-size effects.
Let us consider, for example, the  lines  with $N_0=400$,
shown in Figs.~\ref{general}(b), \ref{general}(d), and \ref{general}(f) for the phase diagram point $\widetilde{n}=1.5,~\widetilde{T}=1.5$.\footnote{The value $N_0=400$ corresponds approximately  to the total number of nucleon participants in most central heavy-ion collisions.}
To have $\omega$,  $\sk$, and $\kurt$ deviate from their thermodynamic limits  by no more than $10\%$ one has to take, respectively, $x\gtrsim0.05$, $x\gtrsim0.10$, and $x\gtrsim0.15$.
This numerical example as well as the general trend of the data presented in Fig.~\ref{general} demonstrate an important conclusion:
For the same $(\widetilde{n},\widetilde{T})$ point, the proximity to the thermodynamic limit behavior is different for different fluctuation measures. 
To reach the same proximity for the higher moments of particle number distribution one needs a larger system (larger $N_0$) and/or larger experimental acceptance (larger $x$).

The finite size effects become stronger in the vicinity of the CP.
For example, the results at  
$\widetilde{n}=1$, $\widetilde{T}=1.2$ are shown in Figs.~\ref{general}(a), \ref{general}(c), and \ref{general}(e).
This $\widetilde{n},\widetilde{T}$ point
is closer to the CP. To reach the proximity to the thermodynamic limits shown by red lines the values of $N_0$ and/or $x$ must be higher than those in Figs.~\ref{general}(b), \ref{general}(d), and \ref{general}(f).

In practice, i.e. in the scenario of a nucleus-nucleus collision, it is not easy to exactly ascertain whether a system is in the thermodynamic limit. However, the model-independent Eqs.~(\ref{omega-tdl})--(\ref{kurt-tdl}) provide a way to estimate to what extent the thermodynamic limit is reached by studying the acceptance dependence of cumulant ratios of a conserved charge. Namely, if at a given system size in some $x$-interval $\omega$ and $\sk$ exhibit a linear decrease with $x$ and $\kurt$ exhibits a parabolic $x$-dependence, the system may be close to the thermodynamic limit. Then, one can use Eqs.~(\ref{omega-tdl})-(\ref{kurt-tdl}) to extract the corresponding GCE values, $\omega_{\rm gce}$, $\sk_{\rm gce}$, and $\kurt_{\rm gce}$.
Also, as larger systems are closer to the thermodynamic limit, it is preferable to study the most central collisions of heavy ions.

When $x$ is close to 0 or 1 some
``oscillations" in the $x$-dependence are visible for moderate values of $N_0$.  
This is connected to the excluded volume restrictions when only few finite-sized particles can fit in the volume $V$.
We explore the finite size effects specifically in Sec.~\ref{sec-V}.

\subsection{Finite size effects} \label{sec-V}
In this subsection we discuss the
thermodynamic limit, $V_0\rightarrow\infty$, for  
finite values of volume $V$.
This corresponds to $x=V/V_0\rightarrow0$ as $V_0\rightarrow\infty$ {\it simultaneously}.
In this case, the free energy, $F=-\,T\,{\rm ln}~Z_{\rm ce}$, of the reservoir with $N_0-N$ particles in the volume $V_0-V$ can be written as
\eq{\nonumber
& F\left(V_0-V,N_0-N,T\right) \cong  
F\left(V_0-V,N_0,T\right)\\
& -\left(\der{F}{N_0}\right)_{V_0,T} N~. \label{F}
}
The partition function (\ref{partition2}) can then be expressed as 
\eq{\label{partition-gce}
Z(V,T)~& = \sum_{N=0}^{N_{\rm max}}Z(N;V,T) 
= Z_{\rm ce}\left(V_0-V,N_0,T\right) \nonumber \\
& \quad \times~\sum_{N=0}^{N_{\rm max}} \exp\left(\frac{\mu_0 N}{T}\right)~Z_{\rm ce}(V,N,T)~, 
}
where  $\left(\partial{F}/\partial{N_0}\right)_{V_0,T}=\mu_0$, is the chemical potential of the reservoir and $N_{\rm max}=\lfloor V/b\rfloor$.
Equation~(\ref{partition-gce}) 
includes the finite size effects because of the finite value of $N_{\rm max}$. This finite size restriction is not very important at the regions of the phase diagram located far away from the CP. It is, however, crucial at the CP when the intensive fluctuation measures  become divergent.  
In the thermodynamic limit $V\rightarrow \infty$, Eq.~(\ref{partition-gce}) leads to $N_{\rm max}\rightarrow \infty$ and the $N$-fluctuation measures in the subensemble approach their GCE values. Their behavior was discussed in Sec.~\ref{sec-vdW}.

In the  vdW model, $\mu_0$ is calculated  as \cite{Vovchenko:2015xja}:
\eq{
\mu_0=-T~{\rm ln}\frac{(V_0-b N_0)\phi(T)}{N_0}+\frac{b\,N_0 T}{V_0-b N_0}-2 a \frac{N_0}{V_0}.
}
The probability distribution (\ref{probability}) is then calculated as
\eq{
W(N;V,T)~=~\frac{A(N;V,T)}{\sum_{N=0}^{N_{\rm max}}A(N;V,T)}~, 
}
where
\eq{\nonumber
A~& =\frac{(N_{\rm max} -N)^N}{N!}\left(\frac{3-\widetilde{n}}{\widetilde{n}}\right)^{-N}\\
& \quad \times~\exp\left[\frac{1}{3-\widetilde{n}}-\frac{9}{4}\frac{1}{\widetilde{T}}\left(1- \frac{3N^2}{2N_{\rm max}}\right) \right]   ~.
\label{AN}
}

\begin{figure}
\includegraphics[width=.49\textwidth]{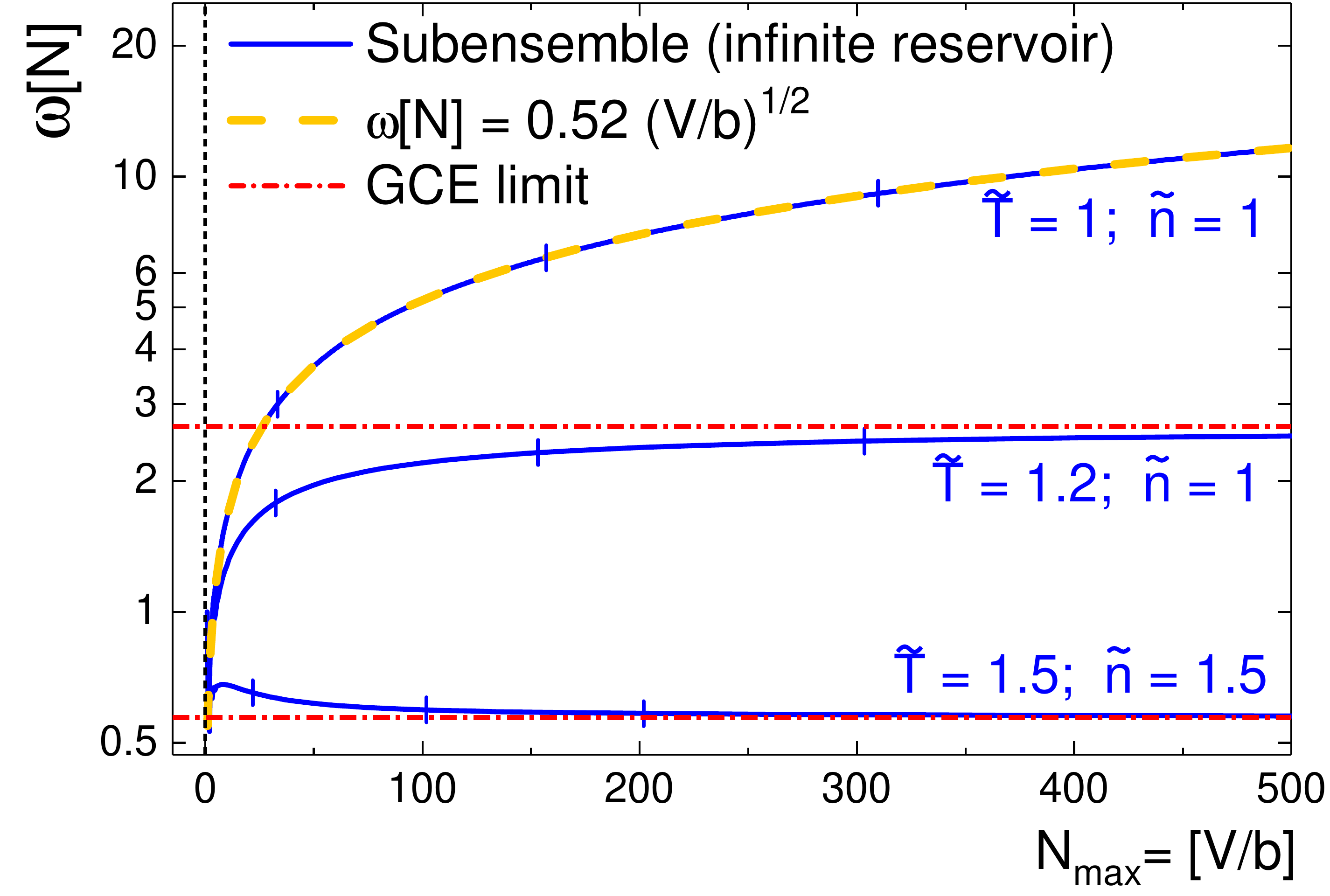}
\caption{\label{fig-fse}
The scaled variance $\omega$ at different $(\widetilde{n},\widetilde{T})$ points in the phase diagram as a function of $N_{\rm max}=V/b$ calculated within the subensemble for the infinite reservoir case, $V_0\rightarrow \infty$. 
The three vertical ticks at each of the lines correspond to values of $\mean{N}$ being $10$, $50$, and $100$.
The scaled variance exhibits a scaling $\omega \sim V^{1/2}$ at the critical point, $\widetilde{n}=1$ and $\widetilde{T}=1$, depicted by the dashed yellow line.
}
 \end{figure}

This agrees with the result of Ref.~\cite{Bzdak:2018uhv} where particle number distributions for finite vdW system were calculated in the ($T$,$\mu$) plane.
The results for the scaled variance $\omega$ in the subensemble with the probability distribution (\ref{AN}) are presented in Fig.~\ref{fig-fse}.
The finite size effects disappear with increasing $N_{\rm max}=V/b$. To approach the GCE limit with the same accuracy larger $V$-values are needed the closer $\widetilde{n}$ and $\widetilde{T}$ are to their critical values. At the CP point $\widetilde{n}=1$ and $\widetilde{T}=1$ the scaled variance in the subensemble behaves as $\omega \sim \sqrt{V}$ and is divergent at $V\rightarrow\infty$.

 \begin{figure}
\includegraphics[width=.49\textwidth]{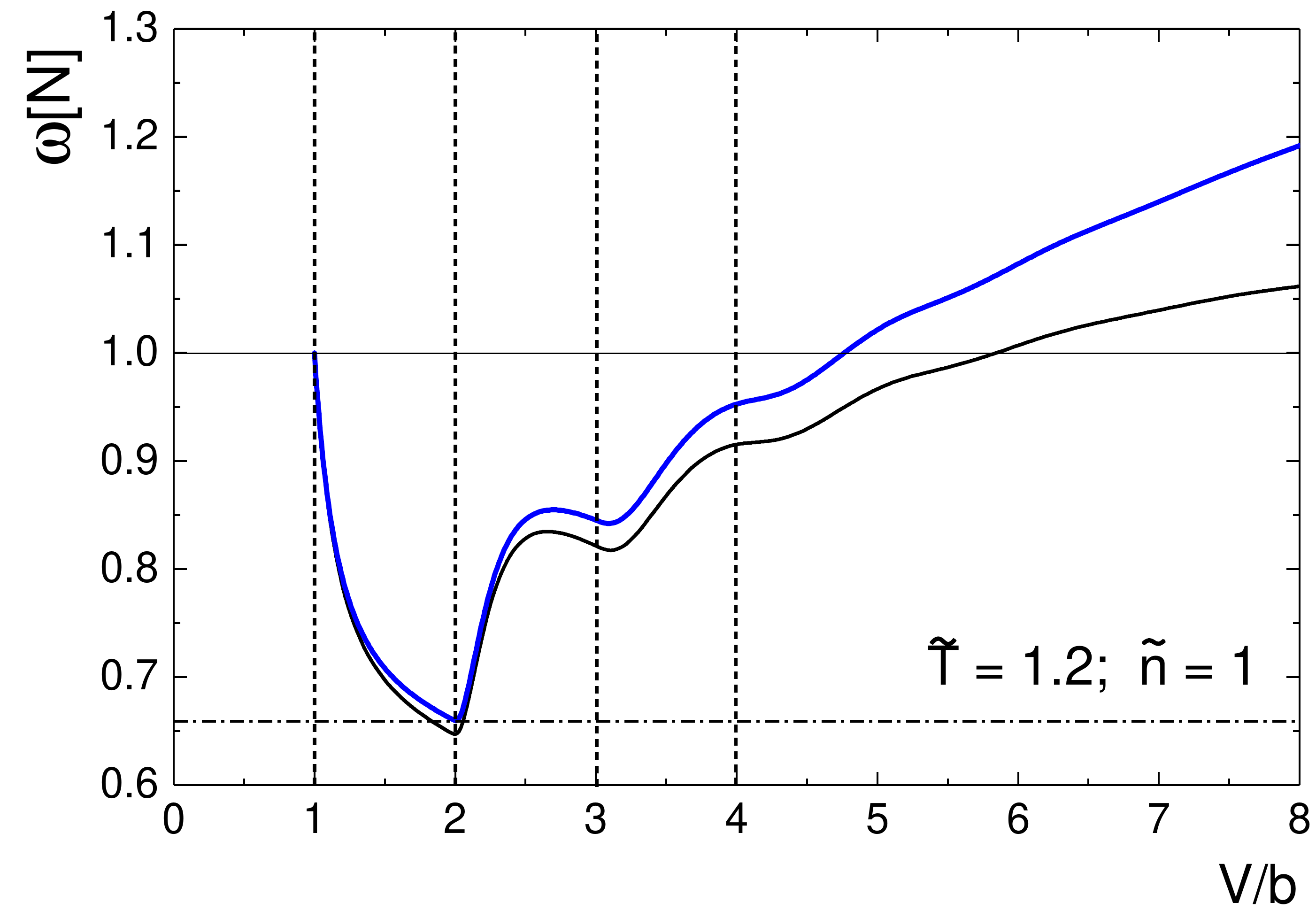}
\caption{\label{fig-ev}
The scaled variance   $\omega$ in the subensemble 
as a function of $V/b$.
Vertical lines show the threshold values of $V/b$. 
Black line corresponds to $N_0=10$ while blue line represents thermodynamic limit $V_0\rightarrow\infty$. 
Cases with $N_0\gtrsim50$ are well approximated by the blue line.
The ``oscillatory" behavior of scaled variance is due to the excluded volume threshold effects, see also Figs.~\ref{general}(a)--\ref{general}(b).
}
 \end{figure}

The ``oscillatory" behavior of the fluctuations in the subensemble  at small volumes $V$ is observed in Fig.~\ref{general} at $x\approx0$ and $x\approx1$.
This will be  illustrated now on example of the scaled variance $\omega$.
When the volume $V$  is so small that only one particle can fit in, $b<V<2b$, the partition function (\ref{partition2}) of the subensemble  is a sum of only two terms
with $N=0$ and $N=1$.
In this case, 
one obtains  $\mean{N^k}=\mean{N}$ for $k=1,2$ and, thus,
\eq{\label{fluct-one-particle}
 \omega[N]=1-\mean{N}<1~.
}
At $V/b\rightarrow 1+0$ one finds $\langle N\rangle\rightarrow 0$. 
Thus,  $\omega[N]\rightarrow 1$, which is in agreement with the Poisson distribution.

We demonstrate the excluded volume threshold effects by depicting in Fig.~\ref{fig-ev} the scaled variance $\omega$ as a function of $V/b$ in the region of small $V/b$. 
The vertical dashed lines show the thresholds of the system volume $V/b$ at $1,2,3$, and $4$ particle level. 
One sees that the excluded volume  threshold effects for $\omega[N]$ are substantial at $V/b\lesssim5$ 
which corresponds to $x\lesssim5(bn)/N_0$.
The same oscillatory behavior of $\omega$ due to the excluded volume threshold effects is also seen in Figs.~\ref{general}(a)--\ref{general}(b) where $\omega$ is presented as a function of $x$.

\section{summary}
\label{summary}

We investigated particle number fluctuations in an interacting thermal subsystem, taking into account effects associated with the global conservation of particle number~(conserved charge) and finite system size.
The total number of particles $N_0$ (total conserved charge) in the whole volume is fixed, in analogy to the final state (net) baryons in heavy-ion collisions, and treated in the canonical ensemble.
The fluctuations of particle number $N$ in a subvolume~(acceptance) $V<V_0$ are described by a statistical ensemble which is distinct from both the canonical and grand canonical ensembles.

The specific calculations have been performed for the van der Waals (vdW) equation of state, which contains a first-order phase transition and a critical point. 
The supercritical temperatures have been considered.
Due to the universality of the critical behavior, we expect our results to reflect generic features of fluctuations near a critical point of a first-order phase transition in the presence of global charge conservation effects.

The global charge conservation influences the fluctuations at any finite value of the subvolume fraction $x \equiv V/V_0$.
In the thermodynamic limit, $N_0\rightarrow \infty$, these effects are in agreement with the recently developed subensemble acceptance procedure~\cite{Vovchenko:2020tsr} and thus can be corrected for analytically.

In a more general case of a finite $N_0$ and finite $x$, both the finite size and global charge conservation effects simultaneously influence the fluctuation measures.
The finite size effects at a fixed value of $N_0$ are the smallest at $x = 1/2$, where the two subsystems are both large.
The magnitude of the finite size effects depends on the proximity of the critical point: the closer the system is to the critical point, the larger are the finite size effects.
This can be understood due to the growth of the correlation length and, correspondingly, fluctuations in the vicinity of the critical point, which become comparable to the total system size.

Threshold effects are observed for very small volumes, $V \gtrsim b$, when only few finite-sized particles fit into the volume.
An oscillatory behavior is observed, associated with the opening of new channels at the thresholds.

The following strategy may be adopted for extracting the GCE values of the cumulant ratios in relativistic heavy-ion collisions.

(i) The behavior of $\omega$ and $\sk$ of the  fluctuations of a conserved charge should be analyzed within several different acceptances (which corresponds to different $x$ values). 
If the linear $x$-dependence 
of $\omega$ and $\sk$ is established, it
can be considered as a signal of approaching the thermodynamic limit
(see  Fig.~\ref{general}). 
Linear fits can then be performed to extract $\omega_{\rm gce}$ and $\sk_{\rm gce}$.

(ii) The finite-size effects have a stronger influence on the kurtosis $\kurt$ compared to $\omega$ and $\sk$.
As the finite-size effects are the smallest at $x = 1/2$, it is advisable to measure $\kurt$ in an acceptance as close to $x = 1/2$ as possible.
One can then extract $\kurt_{\rm gce}$ from experimentally measured $\kurt$ and the previously reconstructed $\sk_{\rm gce}$ using Eq.~(\ref{kurt-tdl}).

It should be noted that our analysis is based on an idealized picture of a homogeneous system in statistical equilibrium.
It does not incorporate the various dynamical effects present in  relativistic heavy-ion collision experiments, detector limitations, as well as system volume, $V_0$, fluctuations.
Moreover, measurements in heavy-ion experiments are performed in the momentum space rather than in the coordinate space. 
The degree of correlation between
momenta and coordinates of particles at freeze-out depends on the collective flow, for example, the longitudinal flow.
To reduce the effects of $V_0$ fluctuations 
the so-called strongly intensive fluctuation measures~\cite{Gorenstein:2011vq,Sangaline:2015bma} may be used.
In future works we plan to include the influence of  dynamical effects, the analysis of strongly intensive fluctuation measures, as well as to  address the connection between the system's separation in coordinate space with the corresponding separation in the momentum space.
We also plan to extend our approach to fully relativistic systems with multiple conserved charges~\cite{Vovchenko:2020gne}, as is appropriate for relativistic heavy-ion collisions.

\vspace{0.5cm}
\section*{Acknowledgments}
The authors are thankful to Marek Gazdzicki, Volker Koch, Carsten Greiner, and Anar Rustamov for fruitful discussions.
This work is partially supported 
by the Target Program of Fundamental Research of the Department of Physics and Astronomy of the National Academy of Sciences of Ukraine
(N~0120U100857).
R.P. and K.T. acknowledge the generous support by the Stiftung Polytechnische Gesellschaft Frankfurt.
V.V. was supported by the
Feodor Lynen program of the Alexander von Humboldt
foundation and by the U.S. Department of Energy, 
Office of Science, Office of Nuclear Physics, under Contract No. DE-AC02-05CH11231.
L.S. thanks the support of the Frankfurt Institute for Advanced
Studies.
J.S. thanks the Samson AG and the BMBF through the ErUM-Data project for funding. This work was supported by the DAAD through a PPP exchange grant. Computational resources were provided by the Frankfurt Center for Scientific Computing (Goethe-HLR).
H.St. acknowledges the support through the Judah M. Eisenberg Laureatus Chair by Goethe University  and the Walter Greiner Gesellschaft, Frankfurt.

\bibliography{F-4.bbl}

\end{document}